\documentclass{aip-cp}

\usepackage{bm}
\usepackage[numbers]{natbib}
\usepackage{graphicx}
\usepackage[utf8]{inputenc}

\begin{document}

\title{Applications Of The Divergence Theorem In Bayesian Inference And MaxEnt}

\author[aff1]{Sergio Davis\corref{cor1}}
\author[aff2]{Gonzalo Gutiérrez}
\eaddress{gonzalo@fisica.ciencias.uchile.cl}

\affil[aff1]{Comisión Chilena de Energía Nuclear, Casilla 188-D, Santiago, Chile.}
\affil[aff2]{Grupo de Nanomateriales, Departamento de Física, Facultad de Ciencias, Universidad de Chile.}
\corresp[cor1]{Corresponding author: sdavis@cchen.cl}

\maketitle

\begin{abstract}
Given a probability density $P({\bf x}|{\boldsymbol \lambda})$, where $\bf x$ represents continuous 
degrees of freedom and $\lambda$ a set of parameters, it is possible to construct a general 
identity relating expectations of observable quantities, which is a generalization of the 
equipartition theorem in Thermodynamics.

In this work we explore some of the consequences of this relation, both in the
context of sampling distributions and Bayesian posteriors, and how it can be
used to extract some information without the need for explicit calculation of
the partition function (or the Bayesian evidence, in the case of posterior
expectations). Together with the general family of fluctuation theorems 
it constitutes a powerful tool for Bayesian/MaxEnt problems.
\end{abstract}

\section{INTRODUCTION}

In Statistical Thermodynamics, a system composed of $N$ particles and connected to a thermal 
bath at temperature $T$ is described by the \emph{canonical ensemble},

\begin{equation}
P({\bf \Gamma}|\beta) = \frac{1}{Z(\beta)}\exp(-\beta \mathcal{H}({\bf \Gamma})),
\label{eq_canonical}
\end{equation}
where $\beta=1/k_B T$, the vector ${\bf \Gamma}=(x_1, x_2, \ldots, x_{3N}, p_1, p_2, \ldots, p_{3N})$
represents the $6N$ degrees of freedom of the system and $\mathcal{H}$ is the
Hamiltonian function, usually of the form

\begin{equation}
\mathcal{H}({\bf \Gamma}) = \sum_{i=1}^N \frac{{{\bf p}_i}^2}{2m_i} + \Phi(x_1, \ldots, x_{3N}).
\end{equation}

Here $\beta$ plays two roles: on the one hand it is the Lagrange multiplier which 
describes the probability distribution; on the other, it is associated with the 
average kinetic energy of the particles via the \emph{equipartition theorem}, 

\begin{equation}
\frac{3Nk_B T}{2} = \Big<\sum_{i=1}^N \frac{{{\bf p}_i}^2}{2m_i}\Big>_\beta
\end{equation}
where the average $\big<\cdot\big>_\beta$ is taken using the canonical distribution in 
Eq. \ref{eq_canonical}. This equipartition theorem, in a slightly more general form~\cite{Huang1987},

\begin{equation}
\beta\Big<\Gamma_j\frac{\partial \mathcal{H}}{\partial \Gamma_i}\Big>_\beta = \delta_{ij},
\end{equation}
is obtained using the divergence (Gauss) theorem, 

\begin{equation}
\int_\Sigma d\Sigma (u {\bf v}\cdot {\bf \hat{n}}) = \int_V dV \left(u\nabla\cdot {\bf v} + {\bf v}\cdot \nabla u\right),
\label{eq_divergence}
\end{equation}
where $V$ is a volume in $\mathbb{R}^{6N}$ representing a region of phase-space, $\Sigma$ is a surface acting as a
boundary of the volume $V$ and $\hat{n}$ is the unit vector normal to $\Sigma$.

\vspace{10pt}

Now, since the seminal paper by Jaynes~\cite{Jaynes1957} we know that Statistical Mechanics and the 
rule for constructing the canonical ensemble is just a particular application of the maximum entropy
(MaxEnt) principle to physics; this suggests that some general form of the equipartition theorem 
holds for any continuous MaxEnt model fulfilling the conditions of the divergence theorem. 

From now on we will work in general terms with a system described by $N$
continuous degrees of freedom denoted collectively by ${\bf x}=(x_1, \ldots,
x_N)$. This system needs not be a ``physical'' system as in the case of the
canonical ensemble before, but can be any set of $N$ unknown quantities over
which we do maximum entropy inference. We seek the optimal probability distribution consistent 
with $m$ expectation constraints

\begin{equation}
\big<f_j({\bf x})\big> = F_j,
\end{equation}
when starting from the \emph{prior}\footnote{Jaynes, in the original formulation
of MaxEnt, takes $P_0$ to be the invariant measure of the state space. However,
recently~\cite{Caticha2006} it has been shown that this is too restrictive: MaxEnt
can be formulated as a method of updating an arbitrary \emph{a priori} distribution under constraints.}
distribution $P_0({\bf x})$. The solution to this problem is the MaxEnt probability

\begin{equation}
P({\bf x}|{\boldsymbol \lambda}) = \frac{1}{Z({\boldsymbol \lambda})}P_0({\bf x})\exp\left(-\sum_{j=1}^m\lambda_j f_j({\bf x})\right).
\label{eq_maxent}
\end{equation}
 
For this probability the authors have recently presented a generalization of the equipartition theorem~\cite{Davis2012},

\begin{equation}
\Big<\frac{\partial}{\partial x_i}\omega({\bf x})\Big>_{\boldsymbol \lambda} + 
\Big<\omega({\bf x})\frac{\partial}{\partial x_i}\ln P_0({\bf x})\Big>_{\boldsymbol \lambda} = 
\sum_{j=1}^m\lambda_j\Big<\omega({\bf x})\frac{\partial}{\partial x_i} f_j({\bf x})\Big>_{\boldsymbol \lambda}.
\label{eq_cvt_maxent}
\end{equation}
where $\omega({\bf x})$ is an arbitrary, differentiable function of the coordinates ${\bf x}$.

This theorem relates the constraining functions $f_j$ with their
conjugate\footnote{Conjugate in the sense of Thermodynamics.} Lagrange multipliers $\lambda_j$ and was thus named the
\emph{conjugate variables theorem}, a neutral name which attempts to avoid any
connection with the physical idea of equipartition (which might also be confused
with the idea of ``asymptotic equipartition'' already present in Information Theory\cite{CoverThomas2006}).

In this work we start by providing a short proof of the CVT, which is more clear and general 
than the one in Ref. \cite{Davis2012}, and then review some applications to Statistical Mechanics, 
dynamical systems and Bayesian inference.

\subsection{Variation inside the expectation: Conjugate variables theorem}

Consider a model $P({\bf x}|{\boldsymbol \lambda})$ where ${\bf x}=(x_1,\ldots,x_N)$ represents a vector
of random (or unknown) continuous variables which describe a system, and
${\boldsymbol \lambda}=(\lambda_1,\ldots,\lambda_m)$ denotes the vector of parameters of the
model. We take every state ${\bf x}$ as a point in $\mathbb{R}^N$.

We construct the expectation of a partial derivative $\partial \omega/\partial x_i$ of
an arbitrary function $\omega({\bf x},{\boldsymbol \lambda})$ of the state and the parameters, and 
rewrite it using the divergence theorem (Equation \ref{eq_divergence}),

\begin{equation}
\Big<\frac{\partial \omega}{\partial x_i}\Big>_{\boldsymbol \lambda} = \int
d{\bf x} P({\bf x}|{\boldsymbol \lambda})\frac{\partial}{\partial
x_i}\omega({\bf x},{\boldsymbol \lambda}) = \int d\Sigma ({\bf \hat{e}}_i \cdot
{\bf \hat{n}})P({\bf x}|{\boldsymbol \lambda})\omega({\bf x},{\boldsymbol
\lambda}) - \int d{\bf x} \omega({\bf x},{\boldsymbol \lambda})\frac{\partial}{\partial x_i}P({\bf x}|{\boldsymbol \lambda})
\end{equation}
where we have chosen $u=P({\bf x}|{\boldsymbol \lambda})$ and ${\bf v}={\bf \hat{e}}_i\omega({\bf x},{\boldsymbol \lambda})$. 
As $P({\bf x}|{\boldsymbol \lambda})$ must be properly normalized, it must vanish at $x_i=\pm\infty$ so the 
surface term also vanishes. Then, the second term can also be written as an expectation, and we finally obtain the identity

\begin{equation}
\Big<\frac{\partial \omega}{\partial x_i}\Big>_{\boldsymbol \lambda} =
-\Big<\omega\frac{\partial}{\partial x_i} \ln P({\bf x}|{\boldsymbol \lambda})\Big>_{\boldsymbol \lambda},
\label{eq_cvt}
\end{equation}
valid for an arbitrary, differentiable function $\omega$. This is the most general form of the 
CVT for any continuous model $P({\bf x}|{\boldsymbol \lambda})$. The particular
case in Eq. \ref{eq_cvt_maxent} is obtained by replacing the MaxEnt probability distribution given
by Eq. \ref{eq_maxent}.

A direct consequence of this identity is readily seen by setting $\omega=1$, it
shows that the condition for the probability maximum is also valid in
expectation over all states, that is,

\begin{equation}
\Big<\frac{\partial}{\partial x_i} \ln P({\bf x}|{\boldsymbol \lambda})\Big>_{\boldsymbol \lambda} = 0.
\label{eq_maximum_exp}
\end{equation}

\subsection{Variation outside the expectation: Fluctuation theorems}

Under the same conditions of the CVT, let us now take the derivative of the expectation
$\Big<\omega({\bf x},{\boldsymbol \lambda})\Big>_{\boldsymbol \lambda}$ with respect to one of 
the parameters, $\lambda_j$,

\begin{equation}
\frac{\partial}{\partial \lambda_j}\Big<\omega\Big>_{\boldsymbol \lambda} = 
\frac{\partial}{\partial \lambda_j}\left(\int d{\bf x}\omega({\bf x},{\boldsymbol \lambda})P({\bf x}|{\boldsymbol \lambda})\right) = 
\int d{\bf x} \left(P({\bf x}|{\boldsymbol \lambda})\frac{\partial \omega}{\partial \lambda_j} 
+ \omega\frac{\partial}{\partial \lambda_j} P({\bf x}|{\boldsymbol \lambda})\right).
\end{equation}

Again, rewriting the second term as an expectation, we have

\begin{equation}
\frac{\partial}{\partial \lambda_j}\Big<\omega\Big>_{\boldsymbol \lambda} =
\Big<\frac{\partial \omega}{\partial \lambda_j}\Big>_{\boldsymbol \lambda} 
+ \Big<\omega\frac{\partial}{\partial \lambda_j}\ln P({\bf x}|{\boldsymbol
\lambda})\Big>_{\boldsymbol \lambda},
\label{eq_fdt}
\end{equation}
which is similar to the CVT in that we have a ``free'' function $\omega(x,\lambda)$. 
This is the most general version of a family of ``fluctuation theorems''. The particular 
case where $P({\bf x}|{\boldsymbol \lambda})$ is a Maximum Entropy model was already 
given by Jaynes in his book\cite{Jaynes2003},

\begin{equation}
\frac{\partial}{\partial \lambda_j}\Big<\omega\Big>_{\boldsymbol \lambda} = 
-\Big<\delta \omega\delta f_j\Big>_{\boldsymbol \lambda}.
\label{eq_fdt_maxent}
\end{equation}
 
Here the right-hand side is in fact a covariance or correlation between
fluctuations. In particular, in the canonical ensemble the following relation
connecting the energy fluctuations with the heat capacity~\cite{Callen1985},

\begin{equation}
\Big<(\mathcal{H}-E)^2\Big>_\beta = -\frac{\partial E}{\partial \beta} = N k_B T^2 c_v,
\end{equation}
is a particular case of Eq. \ref{eq_fdt_maxent} with $f_j=\mathcal{H}=\omega$ and $\lambda_j=\beta$.

Notice that the FDT is also valid in the case of discrete states but continuous
parameters. For instance, in the case of the Poisson distribution,

\begin{equation}
P(k|\lambda) = \frac{\lambda^k \exp(-\lambda)}{k!},
\end{equation}
it is possible to write the FDT as

\begin{equation}
\frac{\partial}{\partial \lambda}\Big<\omega(n, \lambda)\big>_\lambda = 
\Big<\frac{\partial \omega}{\partial \lambda}\Big>_\lambda +
\Big<\omega\left(1-\frac{k}{\lambda}\right)\Big>_\lambda,
\end{equation}
from which it quickly follows (using $\omega=\lambda$) that $\lambda=\big<k\big>_\lambda$.

\newpage

\subsection{Direct applications of the CVT in Thermodynamics}

We may consider the equipartition theorem as the original application of the
CVT, in much the same way as the canonical ensemble was the original application
of MaxEnt. For a classical system with Hamiltonian $\mathcal{H}({\bf \Gamma})$ in the
canonical ensemble, the CVT takes the form 

\begin{equation}
\Big<\frac{\partial \omega}{\partial \Gamma_i}\Big>_\beta =
\beta\Big<\omega\frac{\partial \mathcal{H}}{\partial \Gamma_i}\Big>_\beta
\label{eq_cvt_canonical}
\end{equation}
and the equipartition theorem~\cite{Huang1987} follows by choosing $\omega(\Gamma)=\Gamma_j$,

\begin{equation}
\beta\Big<\Gamma_j\frac{\partial \mathcal{H}}{\partial \Gamma_i}\Big>_\beta = \delta_{ij}.
\end{equation}

From this identity it follows that, if we choose $\Gamma_j$ as any quadratic degree of freedom, i.e., 

\begin{equation}
\mathcal{H} = \alpha_j {\Gamma_j}^2 + \mathcal{H}'(\Gamma_1,\ldots,\Gamma_{j-1},\Gamma_{j+1},\ldots,\Gamma_N)
\end{equation}
then $\Gamma_j\partial_j \mathcal{H}=2\alpha_j {\Gamma_j}^2$ and therefore
$\big<\alpha_j {\Gamma_j}^2\big>_\beta = k_B T/2$, which is the standard
statement of the equipartition theorem, 

\begin{quote}
``Each quadratic degree of freedom in the Hamiltonian contributes exactly $k_BT/2$ to the average energy''.
\end{quote}

A generalization is the following expression for the inverse temperature, first reported by Rugh~\cite{Rugh1997} and
Rickayzen~\cite{Rickayzen2001} in the context of the microcanonical ensemble,

\begin{equation}
\beta = \Big<\nabla\cdot\left[\frac{{\bf v}({\bf \Gamma})}{{\bf v}({\bf \Gamma})\cdot \nabla \mathcal{H}}\right]\Big>_\beta.
\end{equation}
which can be obtained from the vector form of the Equation
\ref{eq_cvt_canonical},

\begin{equation}
\Big<\nabla \cdot {\boldsymbol \omega}\Big>_\beta = \beta\Big<{\boldsymbol \omega}\cdot \nabla \mathcal{H}\Big>_\beta
\end{equation}
by choosing ${\boldsymbol \omega}={\bf v}/({\bf v}\cdot\nabla \mathcal{H})$. The reason this result works for
the canonical ensemble even when the original derivation was in the microcanonical ensemble is, besides the ensemble equivalence in the
thermodynamic limit, the fact that this construction can be extended to any ensemble in Statistical Mechanics 
where the probability distribution of the microstates $P({\bf \Gamma}|S)=\rho(\mathcal{H}({\bf \Gamma}))$ is dependent only on $\mathcal{H}$.

In order to see this, we use the chain rule,

\begin{equation}
\frac{\partial}{\partial \Gamma_i}\ln \rho(\mathcal{H}) = 
\left[\frac{\partial}{\partial E}\ln \rho(E)\Big|_{E=\mathcal{H}}\right]\frac{\partial\mathcal{H}}{\partial \Gamma_i}
\end{equation}
and substitute in Equation \ref{eq_cvt}, leading to 

\begin{equation}
\Big<\frac{\partial \omega}{\partial \Gamma_i}\Big>_S = 
\Big<\hat{\beta}(\mathcal{H})\omega\frac{\partial \mathcal{H}}{\partial \Gamma_i}\Big>_S,
\end{equation}
where we have defined the quantity

\begin{equation}
\hat{\beta}(E) = -\frac{\partial}{\partial E}\ln \rho(E)
\end{equation}
as an analog of the Lagrange multiplier $\beta$. Now the same substitution
${\boldsymbol \omega}={\bf v}/({\bf v}\cdot \nabla \mathcal{H})$ as before produces

\begin{equation}
\Big<\hat{\beta}(\mathcal{H})\Big>_S =
\Big<\nabla\cdot\left[\frac{{\bf v}}{{\bf v}\cdot \nabla \mathcal{H}}\right]\Big>_S.
\end{equation}
which shows that the expectation in the right-hand side is independent of the
choice of ${\bf v}$. The canonical ensemble is the trivial case where $\hat{\beta}$
reduces to the constant $\beta$.

Thus, the CVT reveals a family of ``temperature estimators'' which can be used
to measure the inverse temperature $\beta$ in different ensembles.

\newpage

\subsection{Dynamical systems and the functional version of CVT}

Soon after the proposal by Jaynes of the Maximum Entropy principle, it was
realized~\cite{Filyukov1967, Jaynes1980} that the same principle should apply to
inference over dynamical systems. The application to time-dependent systems
became known as the Maximum Caliber\footnote{Because the entropy of paths is
analogous to the cross section of a tube, or the caliber of a gun barrel.} principle.

A brief and simplified formulation of this version of the principle is as follows. Consider an 
unknown trajectory $x(t)$ of a one--dimensional system where we instantly know the joint probability 
density $P(x(t)=X,\dot{x}(t)=V)=P(X,V|t)$ of position and velocity. The most unbiased
probability \textbf{functional} for the possible trajectories $x(t)$ is, according to the Maximum 
Caliber principle~\cite{Gonzalez2014,Davis2015}, the one that maximizes the path
entropy 

\begin{equation}
\mathcal{S}[P] = -\int Dx() P[x()]\ln P[x()]
\end{equation}
subjected to the constraints

\begin{equation}
\Big<\delta(x(t)-X)\delta(\dot{x}(t)-V)\Big> = P(X, V|t).
\end{equation}

The solution is of the form,

\begin{equation}
P[x()|\mu()] = \frac{1}{\eta[\mu()]}\exp\left(-\int dX dV \int_0^\tau dt \mu(X, V,
t)\delta(x(t)-X)\delta(\dot{x}(t)-V)\right),
\end{equation}
which after explicitly integrating the delta functions, gives

\begin{equation}
P[x()|\alpha] = \frac{1}{\eta(\alpha)}\exp\left(-\int_0^\tau dt \mu(x(t), \dot{x}(t), t)\right)
= \frac{1}{\eta(\alpha)}\exp\left(-\frac{1}{\alpha}A[x()]\right),
\end{equation}
where

\begin{equation}
A[x()] = \int_0^\tau dt \mathcal{L}(x(t), \dot{x}(t), t)
\end{equation}
has the form of the action in classical mechanics with Lagrangian
$\mathcal{L}(x,\dot{x}; t)=\alpha\mu(x,\dot{x}, t)$, and we have extracted a global
factor $1/\alpha$ with units of action from the original Lagrange multiplier $\mu$.

By discretizing time, i.e., considering the trajectory $x(t)$ as a sequence of
discrete time snapshots $x=(x_1,\ldots,x_N)$ separated by a time step $\delta t$
so that $\tau=N\Delta t$, we see that the Maximum Caliber problem can be mapped into a MaxEnt 
problem with $N$ degrees of freedom. It follows that, by the identification~\cite{Gelfand2000} 

\begin{equation}
\frac{1}{\Delta t}\frac{\partial}{\partial x_i} \equiv \frac{\delta}{\delta x(t)}
\end{equation}
the CVT takes the \emph{functional} form

\begin{equation}
\Big<\frac{\delta W[x()]}{\delta x(t')}\Big>_\mathcal{I} = \frac{1}{\alpha}\Big<W[x()]\frac{\delta A[x()]}{\delta x(t')}\Big>_\mathcal{I}.
\end{equation}
where now $W$ is any arbitrary, differentiable \emph{functional}. Taking $W=1$ we see
that the Euler-Lagrange equation for the Lagrangian of the problem is fulfilled in
expectation,

\begin{equation}
\Big<\frac{\delta A[x()]}{\delta x(t)}\Big>_\mathcal{I} = \Big<\frac{\partial \mathcal{L}}{\partial x}-\frac{d}{dt}\left(\frac{\partial \mathcal{L}}{\partial \dot{x}}\right)\Big>_\mathcal{I} = 0.
\end{equation}
which is, of course, Equation \ref{eq_maximum_exp}. This result for the particular case of a one-dimensional particle in a potential
was obtained, in discretized form, in Ref. \cite{Gonzalez2014}, and shows that the ensemble of trajectories follows Newton's law of motion,

\begin{equation}
\Big<\frac{dp}{dt}\Big> = \Big<-\frac{d\Phi(x)}{dt}\Big>,
\end{equation}
in expectation, where $p=m\dot{x}$ is the momentum and $\Phi(x)$ the potential energy.

\subsection{CVT in Bayesian parameter estimation}

Consider a statistical model $P(x|\theta)$ for which we know $n$ observations 
$x_1,\ldots,x_n$, denoted collectively by $D$. The posterior
distribution for the parameter $\theta$ is given by Bayes' theorem,

\begin{equation}
P(\theta|D) = \frac{P(\theta|\mathcal{I}_0)P(D|\theta)}{P(D|\mathcal{I}_0)},
\end{equation}
 
For the case where the $n$ observations are statistically independent, we have

\begin{equation}
P(D|\theta) = \prod_{i=1}^n P(x_i|\theta)
\end{equation}
and therefore

\begin{equation}
P(\theta|D) = \frac{1}{\eta}P(\theta|\mathcal{I}_0)\exp\left(\sum_{i=1}^n \ln P(x_i|\theta)\right),
\end{equation}
where $\eta=P(D|\mathcal{I}_0)$. Now the $x_i$ are given (could be taken as
the fixed parameters of the posterior model) and the single variable $\theta$ is the
sole degree of freedom of our problem. The CVT for this posterior distribution is

\begin{equation}
\Big<\frac{d\omega(\theta)}{d\theta}\Big>_D = -\Big<\omega(\theta)\frac{\partial}{\partial \theta}\ln P(\theta|D)\Big>_D 
= -\Big<\omega(\theta)\left\{\frac{\partial}{\partial \theta}\ln P(\theta|\mathcal{I}_0) + n\overline{\frac{\partial}{\partial \theta}\ln
P(x|\theta)}\right\}\Big>_D.
\label{eq_cvt_parestim}
\end{equation}
where $\overline{a}=\frac{1}{n}\sum_{i=1}^n a(x_i)$ is the arithmetic mean over the data $D$. If our original model $P(x|\theta)$ is a MaxEnt model,

\begin{equation}
P(x|\theta) = \frac{1}{Z(\theta)}\exp(-\theta f(x)),
\end{equation} 
then Equation \ref{eq_cvt_parestim} simplifies to 

\begin{equation}
\Big<\frac{d\omega(\theta)}{d\theta}\Big>_D + \Big<\omega(\theta)\frac{\partial}{\partial \theta} \ln P(\theta|\mathcal{I}_0)\Big>_D 
= n\Big<\omega(\theta)\left[\overline{f} - F(\theta)\right]\Big>_D,
\end{equation}
where we have used 

\begin{equation}
-\frac{d}{d\theta} \ln Z(\theta) = \big<f\big>_\theta = F(\theta).
\end{equation}

For a large number of observations $n$, the effect of the prior vanishes and we simply get

\begin{equation}
\Big<\frac{d\omega(\theta)}{d\theta}\Big>_D = n\Big<\omega(\theta)\left[\overline{f} - F(\theta)\right]\Big>_D.
\end{equation}

From this last identity it follows that 

\begin{eqnarray}
\lim_{n \rightarrow \infty} \overline{f} \rightarrow \Big<F(\theta)\Big>_D = \Big<f(x)\Big>_D \\
\Big<\left(\Delta F(\theta)\right)^2\Big>_D = -\frac{1}{n}\Big<\frac{dF(\theta)}{d\theta}\Big>_D
\end{eqnarray}
by using $\omega=1$ and $\omega=\overline{f}-F(\theta)$ respectively. Together
they imply the law of large numbers: if $dF(\theta)/d\theta$ is finite, then $\overline{f}=\big<f(x)\big>_D$ when $n \rightarrow \infty$.
Note that 

\begin{equation}
\frac{dF(\theta)}{d\theta}=-\Big<\left(\Delta f(x)\right)^2\Big>_\theta
\end{equation}
by the theorem in Equation \ref{eq_fdt_maxent}.

\section{CONCLUSIONS}

We have shown the versatility of the divergence theorem applied to
probability densities of continuous systems, in a form we have called the
conjugate variables theorem (CVT). The same relation that provides the meaning of 
temperature associated with average kinetic energy in Statistical Mechanics can 
be used to obtain properties of non-equilibrium systems in terms of path
averages and as a shortcut in Bayesian parameter estimation. The conjugate
variables theorem (Eq. \ref{eq_cvt}) together with the fluctuation theorem in 
Eq. \ref{eq_fdt}, provide a new set of tools to approach inference problems for continuous variables/parameters.

\section{ACKNOWLEDGMENTS}

SD acknowledges support from FONDECYT grant 1140514.


\bibliographystyle{aipnum-cp}
\bibliography{cvt}

\begin{thebibliography}{14}%
\makeatletter
\providecommand \@ifxundefined [1]{%
 \@ifx{#1\undefined}
}%
\providecommand \@ifnum [1]{%
 \ifnum #1\expandafter \@firstoftwo
 \else \expandafter \@secondoftwo
 \fi
}%
\providecommand \@ifx [1]{%
 \ifx #1\expandafter \@firstoftwo
 \else \expandafter \@secondoftwo
 \fi
}%
\providecommand \natexlab [1]{#1}%
\providecommand \enquote  [1]{``#1''}%
\providecommand \bibnamefont  [1]{#1}%
\providecommand \bibfnamefont [1]{#1}%
\providecommand \citenamefont [1]{#1}%
\providecommand \href@noop [0]{\@secondoftwo}%
\providecommand \href [0]{\begingroup \@sanitize@url \@href}%
\providecommand \@href[1]{\@@startlink{#1}\@@href}%
\providecommand \@@href[1]{\endgroup#1\@@endlink}%
\providecommand \@sanitize@url [0]{\catcode `\\12\catcode `\$12\catcode
  `\&12\catcode `\#12\catcode `\^12\catcode `\_12\catcode `\%12\relax}%
\providecommand \@@startlink[1]{}%
\providecommand \@@endlink[0]{}%
\providecommand \url  [0]{\begingroup\@sanitize@url \@url }%
\providecommand \@url [1]{\endgroup\@href {#1}{\urlprefix }}%
\providecommand \urlprefix  [0]{URL }%
\providecommand \Eprint [0]{\href }%
\providecommand \doibase [0]{http://dx.doi.org/}%
\providecommand \selectlanguage [0]{\@gobble}%
\providecommand \bibinfo  [0]{\@secondoftwo}%
\providecommand \bibfield  [0]{\@secondoftwo}%
\providecommand \translation [1]{[#1]}%
\providecommand \BibitemOpen [0]{}%
\providecommand \bibitemStop [0]{}%
\providecommand \bibitemNoStop [0]{.\EOS\space}%
\providecommand \EOS [0]{\spacefactor3000\relax}%
\providecommand \BibitemShut  [1]{\csname bibitem#1\endcsname}%
\let\auto@bib@innerbib\@empty
\bibitem [{\citenamefont {Huang}(1987)}]{Huang1987}%
  \BibitemOpen
  \bibfield  {author} {\bibinfo {author} {\bibfnamefont {K.}~\bibnamefont
  {Huang}},\ }\href@noop {} {\emph {\bibinfo {title} {Statistical Mechanics}}}\
  (\bibinfo  {publisher} {Wiley},\ \bibinfo {year} {1987})\BibitemShut
  {NoStop}%
\bibitem [{\citenamefont {Jaynes}(1957)}]{Jaynes1957}%
  \BibitemOpen
  \bibfield  {author} {\bibinfo {author} {\bibfnamefont {E.~T.}\ \bibnamefont
  {Jaynes}},\ }\href@noop {} {\bibfield  {journal} {\bibinfo  {journal}
  {Physical Review}\ }\textbf {\bibinfo {volume} {106}},\ \unskip\ \bibinfo
  {pages} {620--630} (\bibinfo {year} {1957})}\BibitemShut {NoStop}%
\bibitem [{\citenamefont {Caticha}\ and\ \citenamefont
  {Giffin}(2006)}]{Caticha2006}%
  \BibitemOpen
  \bibfield  {author} {\bibinfo {author} {\bibfnamefont {A.}~\bibnamefont
  {Caticha}}\ and\ \bibinfo {author} {\bibfnamefont {A.}~\bibnamefont
  {Giffin}},\ }\href@noop {} {\bibfield  {journal} {\bibinfo  {journal} {AIP
  Conf. Proc.}\ }\textbf {\bibinfo {volume} {872}},\ p.~\bibinfo {pages} {31}
  (\bibinfo {year} {2006})}\BibitemShut {NoStop}%
\bibitem [{\citenamefont {Davis}\ and\ \citenamefont
  {Guti\'errez}(2012)}]{Davis2012}%
  \BibitemOpen
  \bibfield  {author} {\bibinfo {author} {\bibfnamefont {S.}~\bibnamefont
  {Davis}}\ and\ \bibinfo {author} {\bibfnamefont {G.}~\bibnamefont
  {Guti\'errez}},\ }\href@noop {} {\bibfield  {journal} {\bibinfo  {journal}
  {Phys. Rev. E}\ }\textbf {\bibinfo {volume} {86}},\ p.\ \bibinfo {pages}
  {051136} (\bibinfo {year} {2012})}\BibitemShut {NoStop}%
\bibitem [{\citenamefont {Cover}\ and\ \citenamefont
  {Thomas}(2006)}]{CoverThomas2006}%
  \BibitemOpen
  \bibfield  {author} {\bibinfo {author} {\bibfnamefont {T.~M.}\ \bibnamefont
  {Cover}}\ and\ \bibinfo {author} {\bibfnamefont {J.~A.}\ \bibnamefont
  {Thomas}},\ }\href@noop {} {\emph {\bibinfo {title} {Elements of Information
  Theory}}}\ (\bibinfo  {publisher} {John Wiley and Sons},\ \bibinfo {year}
  {2006})\BibitemShut {NoStop}%
\bibitem [{\citenamefont {Jaynes}(2003)}]{Jaynes2003}%
  \BibitemOpen
  \bibfield  {author} {\bibinfo {author} {\bibfnamefont {E.~T.}\ \bibnamefont
  {Jaynes}},\ }\href@noop {} {\emph {\bibinfo {title} {Probability Theory: The
  Logic of Science}}}\ (\bibinfo  {publisher} {Cambridge University Press},\
  \bibinfo {year} {2003})\BibitemShut {NoStop}%
\bibitem [{\citenamefont {Callen}(1985)}]{Callen1985}%
  \BibitemOpen
  \bibfield  {author} {\bibinfo {author} {\bibfnamefont {H.}~\bibnamefont
  {Callen}},\ }\href@noop {} {\emph {\bibinfo {title} {Thermodynamics and an
  Introduction to Thermostatistics}}}\ (\bibinfo  {publisher} {Wiley},\
  \bibinfo {year} {1985})\BibitemShut {NoStop}%
\bibitem [{\citenamefont {Rugh}(1997)}]{Rugh1997}%
  \BibitemOpen
  \bibfield  {author} {\bibinfo {author} {\bibfnamefont {H.~H.}\ \bibnamefont
  {Rugh}},\ }\href@noop {} {\bibfield  {journal} {\bibinfo  {journal} {Phys.
  Rev. Lett.}\ }\textbf {\bibinfo {volume} {78}},\ \unskip\ \bibinfo {pages}
  {772--774} (\bibinfo {year} {1997})}\BibitemShut {NoStop}%
\bibitem [{\citenamefont {Rickayzen}\ and\ \citenamefont
  {Powles}(2001)}]{Rickayzen2001}%
  \BibitemOpen
  \bibfield  {author} {\bibinfo {author} {\bibfnamefont {G.}~\bibnamefont
  {Rickayzen}}\ and\ \bibinfo {author} {\bibfnamefont {J.~G.}\ \bibnamefont
  {Powles}},\ }\href@noop {} {\bibfield  {journal} {\bibinfo  {journal} {J.
  Chem. Phys.}\ }\textbf {\bibinfo {volume} {114}},\ p.\ \bibinfo {pages}
  {4333} (\bibinfo {year} {2001})}\BibitemShut {NoStop}%
\bibitem [{\citenamefont {Filyukov}\ and\ \citenamefont
  {Karpov}(1967)}]{Filyukov1967}%
  \BibitemOpen
  \bibfield  {author} {\bibinfo {author} {\bibfnamefont {A.~A.}\ \bibnamefont
  {Filyukov}}\ and\ \bibinfo {author} {\bibfnamefont {V.~Y.}\ \bibnamefont
  {Karpov}},\ }\href@noop {} {\bibfield  {journal} {\bibinfo  {journal}
  {Inzh.-Fiz. Zh.}\ }\textbf {\bibinfo {volume} {13}},\ p.\ \bibinfo {pages}
  {624} (\bibinfo {year} {1967})}\BibitemShut {NoStop}%
\bibitem [{\citenamefont {Jaynes}(1980)}]{Jaynes1980}%
  \BibitemOpen
  \bibfield  {author} {\bibinfo {author} {\bibfnamefont {E.~T.}\ \bibnamefont
  {Jaynes}},\ }\href@noop {} {\bibfield  {journal} {\bibinfo  {journal} {Ann.
  Rev. Phys. Chem.}\ }\textbf {\bibinfo {volume} {31}},\ \unskip\ \bibinfo
  {pages} {579--601} (\bibinfo {year} {1980})}\BibitemShut {NoStop}%
\bibitem [{\citenamefont {Gonz\'alez}, \citenamefont {Davis},\ and\
  \citenamefont {Guti\'errez}(2014)}]{Gonzalez2014}%
  \BibitemOpen
  \bibfield  {author} {\bibinfo {author} {\bibfnamefont {D.}~\bibnamefont
  {Gonz\'alez}}, \bibinfo {author} {\bibfnamefont {S.}~\bibnamefont {Davis}}, \
  and\ \bibinfo {author} {\bibfnamefont {G.}~\bibnamefont {Guti\'errez}},\
  }\href@noop {} {\bibfield  {journal} {\bibinfo  {journal} {Found. Phys.}\
  }\textbf {\bibinfo {volume} {44}},\ p.\ \bibinfo {pages} {923} (\bibinfo
  {year} {2014})}\BibitemShut {NoStop}%
\bibitem [{\citenamefont {Davis}\ and\ \citenamefont
  {González}(2015)}]{Davis2015}%
  \BibitemOpen
  \bibfield  {author} {\bibinfo {author} {\bibfnamefont {S.}~\bibnamefont
  {Davis}}\ and\ \bibinfo {author} {\bibfnamefont {D.}~\bibnamefont
  {González}},\ }\href@noop {} {\bibfield  {journal} {\bibinfo  {journal} {J.
  Phys. A: Math. Theor.}\ }\textbf {\bibinfo {volume} {48}},\ p.\ \bibinfo
  {pages} {425003} (\bibinfo {year} {2015})}\BibitemShut {NoStop}%
\bibitem [{\citenamefont {Gelfand}\ and\ \citenamefont
  {Fomin}(2000)}]{Gelfand2000}%
  \BibitemOpen
  \bibfield  {author} {\bibinfo {author} {\bibfnamefont {I.~M.}\ \bibnamefont
  {Gelfand}}\ and\ \bibinfo {author} {\bibfnamefont {S.~V.}\ \bibnamefont
  {Fomin}},\ }\href@noop {} {\emph {\bibinfo {title} {Calculus of
  Variations}}}\ (\bibinfo  {publisher} {Dover Publications},\ \bibinfo {year}
  {2000})\BibitemShut {NoStop}%
\end{thebibliography}%

\end{document}